\documentclass[a4paper,12pt]{article}
\pdfoutput=1 

\usepackage{jheppub} 

\usepackage[T1]{fontenc} 

\title{{\boldmath Gauge Theory Amplitudes from Cubic Scalar Feynman diagrams }}

\author[a,b]{Roji Pius}

\affiliation[a]{The Institute of Mathematical Sciences,\\IV Cross Road, C.I.T. Campus, Taramani,
Chennai, India-600113}
\affiliation[b]{Homi Bhabha National Institute,\\Training School Complex, Anushakti Nagar, Mumbai,
India 400094}

\emailAdd{rojipius@imsc.res.in}

\abstract{ Feynman diagrams are the foremost tool in the perturbative study of quantum field theory. In gauge theories, the full potential of this tool is revealed when it is combined with the Slavanov-Taylor identities associated with the local gauge symmetry.  Hence, it is desirable to have perturbative expansion of  scattering amplitudes that combine the graphical nature of Feynman diagram expansion and the reations among various diagrams due to the Slavanov-Taylor identities.
  In a remarkable paper,  motivated by the similarity between graph homology and the gauge theory BRST homology, Kreimer, Sars and van Suijlekom found such an expansion.  The implication of this expansion is that the amplitudes in a generic non-abelian gauge theory can be constructed by transmuting the renormalised integrands of trivalent Feynman diagrams in a scalar quantum field theory.  In this paper, we  show that this surprising connection  naturally emerges from Witten's open string field theory in the field theory limit.   }

\begin{document} 
\maketitle
\flushbottom

\section{Introduction} 


Feynman integrals are one of the most important tool in the perturbative study of quantum field theory. Unfortunately, they produce enormous number of tensor integrals for every Feynman diagrams in  non-abelian gauge theories. These tensor integrals make the perturbative analysis of gauge theories an arduous undertaking. However,  the computations can be significantly simplified by relating various Feynman diagrams using  the Slavanov-Taylor idenities associated with the local gauge symmetry  \cite{Slavnov:1972fg, Taylor:1971ff}.  For example,  the renormalizibility of non-abelian Yang-Mills was proved using these identities  \cite{tHooft:1971qjg, Becchi:1975nq}.   Hence, it is desirable to have perturbative expansion of  scattering amplitudes that combine the graphical nature of Feynman diagram expansion and the reations among various diagrams due to the Slavanov-Taylor identities.

 Kreimer, Sars and van Suijlekom in a remarkable work \cite{Kreimer:2012jw} found such an expansion of the scattering amplitudes in non-abelian gauge theories.  This perturbative expansion unravelled a surprising relation between scalar quantum field theory with cubic interaction and quantum non-abelin gauge theory. The scalar field is assumed to be in the adjoint representation of the gauge group of the corresponding  gauge theory.  Using this formalism  the renormalized integrand of a generic non-abelian gauge theory amplitude can be constructed by transmuting the renormalized integrands of scalar field theory Feynmann diagrams with only cubic interactions.  The transmutation is done by applying a special differential of the momenta that flow along the edges of the trivalent Feynman diagram, known as the Corolla differentials, on the renormalized integrands of the scalar field theory.  The Corolla differentials can be systematically constructed using the Corolla polynomials, which are graph polynomials defined with respect to trivalent graphs.  It demonstrates that a covariant quantization of gauge fields is possible without the need of introducing ghost fields. The authors of \cite{Kreimer:2012jw}, uncovered  this relation  by comparing the graph and cycle homology with BRST homology and by realizing an interpretation of Slavnov–Taylor identities through the Corolla diﬀerentials.   
 
 In this paper, we decipher this mysterious relation using Witten's forumation of  open bosonic  string field theory \cite{Witten:1985cc}. The most impressive feature of  this formulation is that the resulting string field theory contains only cubic interaction. Moreover, in the field theory limit it the low energy effective action is  given by a non-abelian gauge theory \cite{Coletti:2003ai}.  By exploring these features, we show that the connection between non-abelian gauge theories and cubic scalar theory naturally emerges from Witten's open string field theory in the field theory limit. We also identify that the action of Corolla differential on the integrand of a trivalent Feynman diagram in a scalar quantum field theory produces the leading term in the $\alpha' \to 0$ limit of the same diagram in Witten's string field theory.  The fact that we are able to derive the Corolla differentials from string field theory hints towards an astounding interplay between graph homology and string field theory. 
 
 The organization of this  paper is as follows. In section \ref{GACD}, we describe the alogirthm for obtaining the gauge theory amplitudes by transmuting the trivalent Feynman diagram integrands of scalar quantum field theory using the Corolla differentials.  In section \ref{offamplitudes}, using Witten's string field theory we express an arbitrary gauge theory amplitude as a sum over trivalent diagrams. In section \ref{CDSFT}, the expansion of gauge theory amplitudes in terms of Corolla differentials and scalar quantum field theory integrands is obtained by taking the field theory limit of the amplitude in Witten's open string field theory describing the scattering of massless gauge particles.  We conclude in section \ref{conclusion} by discussing some of the interesting future directions.  The derivation of cubic and quartic interactions in gauge theory action from string field theory are expalined in the appendices.
 

 \section{Gauge theory amplitudes and Corolla differentials}\label{GACD}
 
 Consider the scattering of $n$ arbitrary physical states in  a pure non-abelian gauge theory. The perturbative expansion of the amplitude $\mathcal{A}_n$ of this process can be organised as a sum over the loops in the Feynaman diagrams associated with the process
 \begin{equation}\label{An}
 \mathcal{F}_n=\sum_{l}\int \prod\limits_{i=1}^l d\xi^{\mu_i}_{i} ~\mathcal{F}_{l,n}\left(\xi_1^{\mu_1},\cdots,\xi_l^{\mu_l} \right)
 \end{equation}
The integrand  $\mathcal{F}_{l,n}$ is obtained by the adding the integrands associated with all the Feynman diagrams $\Gamma_{l,n}$ in the gauge theory having $n$ external legs and $l$ loops
  \begin{equation}\label{An}
\mathcal{F}_{l,n}\left(\xi_1^{\mu_1},\cdots,\xi_l^{\mu_l} \right) =\sum_{\Gamma_{l,n}} \tilde I_{\Gamma_{l,n}}\left(\xi_1^{\mu_1},\cdots,\xi_l^{\mu_l} \right)
 \end{equation}
 where the variables  $\left\{\xi_1^{\mu_1},\cdots,\xi_l^{\mu_l} \right\}$ are the loop momentum flowing along the set of indpendent loops $\left\{C_1,\cdots,C_l \right\}$ in $\Gamma_{l,n}$.
 
  Kreimer, Sars and van Suijlekom  in \cite{Kreimer:2012jw}  showed that the gauge theory integrand  $A_{l,n}$ can be constructed  by transmutating  the  integrands $ I\left(\Upsilon_{l,n}\right)$ of  all the trivalent Feynman diagrams $\Upsilon_{l,n}$ in a  scalar quantum field theory having $n$ external legs and $l$ loops by acting with the Corolla differential $\mathcal{D}\left(\Upsilon_{l,n}\right)$ associated with the graph $\Upsilon_{l,n}$
\begin{equation}\label{PhitoYM}
\mathcal{F}_{l,n}=\sum_{\Upsilon_{l,n}} \mathcal{D}\left(\Upsilon_{l,n}\right) I_{\Upsilon_{l,n}}
\end{equation}
The information about the relations among the Feynman diagrams in a gauge theory due to the Slavanov–Taylor identities are  encoded in the structure of $\mathcal{D}\left(\Upsilon_{l,n}\right)$. Corolla differentials are differentials of the momentum flowing along the edges of a trivalent Feynamn diagram in the scalar quantum field theory. These are constructed from the Corolla polynomials which are defined by  making implicit use of graph homology.  

\subsection{Corolla polynomials}
Let us  briefly describe the definition of the Corolla polynomial associated with a trivalent graph.  An arbitrary trivalent graph $\Upsilon$ that is characterised by a set of vertices $\Upsilon^{[0]}$ and  a set of edges $\Upsilon^{[1]}=\Upsilon^{[1]}_{\text{ext}}\cup \Upsilon^{[1]}_{\text{int}}$ , where $\Upsilon^{[1]}_{\text{ext}}$  are the eternal edges and $\Upsilon^{[1]}_{\text{int}}$ are the internal edges of the graph.  Assume that $v$ denotes a vertex in $\Upsilon$ and $n(v)$ denotes the set of edges that are connected to $v$, i.e. $v\in \Upsilon^[0]$ and $n(v)\in \Upsilon^{[1]}$. Using this data we can introduce the notion of a half-edge. It is defined as the tuple
\begin{equation}\label{halfedge}
h \equiv \left( v,e\right), \qquad v\in \Upsilon^{[0]} \qquad e\in n(v).
\end{equation}
A choice of ordering or direction for all the edges of $\Upsilon$ can be encoded in the incidence matrix $\epsilon\left(\Upsilon\right)$. Its components are defined as 
\begin{align}\label{incidence}
\epsilon_{ve}\left(\Upsilon\right)=\begin{cases}
    +1, & \text{if the vertex $v$ is the end point of the edge $e$}.\\
    -1, & \text{if the vertex $v$ is the starting point of the edge $e$}.\\
    ~~ 0, & \text{if the edge $e$ is not incident on the vertex $v$.}
    \end{cases}
\end{align}
Suppose that $\Upsilon$ contains many loops and all of them can be generated by a set of basis loops denoted by $\mathbb{C}\left(\Upsilon\right)$. The orientation of a loop $C\in \mathbb{C}\left(\Upsilon\right)$ can be specified by fixing the incidence matrix $\epsilon^C\left(\Upsilon\right)$ associated with vertices in $C$. 

In order to construct the Corolla polynomial,  we will associate three variable $a_{h+},a_{h_-}$ and $b_h$. The half-edge $h_+$ denote the half edge that is successor to the half edge  $h$ and te half-edge $h_-$ denote its predecessor according to the orientation of $\Upsilon$.  Then for each vertex $v$ let us  define  
\begin{equation}\label{vertexV}
\mathcal{V}_{v}\equiv\sum\limits_{h\in n(v)}b_h\left(a_{h_-}+a_{h_+}\right).
\end{equation}
Similarly,  for each loop $C_j$ define  
\begin{equation}\label{loopG}
\mathcal{G}_{C_j}\equiv\sum\limits_{q\in\{+,-\}}\prod\limits_{v\in C_j^{[0]}}a_{h\left(C_j,v\right)_q}
\end{equation}
where $h\left(C_j,v\right)$ denotes the half edge of $\Upsilon$ incident on $v$ but not in the loop $C_j$. Then the Corolla polynomial associated with $\Upsilon$ used to constriuct the integrands of pure Yang-Mills theory  amplitude is defined as
\begin{align}\label{corolla}
\mathcal{C}\left(\Upsilon\right) \equiv \sum\limits_{i}^{\infty}\left(-1\right)^i\mathcal{C}^i\left(\Upsilon\right)
\end{align}
The summand $\mathcal{C}^i\left(\Upsilon\right)$ is given by
\begin{align}\label{corollailoop}
\mathcal{C}^i\left(\Upsilon\right)\equiv \sum\limits_{\substack{C_1,\cdots, C_i\in \mathcal{L}\left(\Upsilon\right)\\ C_j \text{ pairwise disjoint}}}\left(\prod\limits_{j=1}^i\mathcal{G}_{C_j}\right)\left(\prod\limits_{\substack{v\in \Upsilon^{[0]}\\v\notin \cup_{k=1}^i C_k^{[0]}}}\mathcal{V}_v\right)
\end{align}
where $\mathcal{L}\left(\Upsilon\right)$ is the set of all loops in $\Upsilon$. Now construct the polynomial $\overline{\mathcal{C}}\left(\Upsilon\right)$ infused with the information about the gauge group and coupling constant of the gauge theory  by multiplying the Corolla polynomial with the color factor and the appropriate power of the coupling constant
\begin{align}\label{corollaNA}
\overline{\mathcal{C}}\left(\Upsilon\right) \equiv \text{i}^{|\Upsilon^{[1]}|}g_{\text{YM}}^{|\Upsilon^{[0]}|} \text{color}\left(\Upsilon\right)\mathcal{C}\left(\Upsilon\right)
\end{align}

\subsection{Corolla differentials}

Let us associate to each edge $e$ in $\Upsilon$ a four momentum $k^{\mu}_e$ and to each half edge $h\equiv \left(v,e \right)$ assign the following momentum vector 
\begin{equation}\label{hedgexi}
\epsilon_{ve}{k'}^{\mu}_e \equiv \epsilon_{ve}{k}^{\mu}_e +\sum_{C\in \mathbb{C}\left(\Gamma\right)}\sum_{e\in C^{[1]}}\epsilon_{ve}^C k^{\mu}_C
\end{equation}
where $k_C^{\mu}$ is the loop momenta flowing along the loop $C$.   At this stage the four momentum flowing through different edges of $\Upsilon$ are not assumed to be related by the momentum conservation at the vertices $v$'s of the graph.  Finally, the Corolla differential for pure non-abelian gauge theory can be obtained by replacing of the variables in Corolla polynomial as follows\cite{Kreimer:2012jw, Laddha:2024qtn, Prinz:2016fka}
\begin{align}\label{CtoD}
\mathcal{D}\left(\Upsilon\right)= \overline{\mathcal{C}}\left(\Upsilon\right)\Bigg|_{\substack{ a_{h_q}\to \mathcal{A}_{h_q}\\ b_h\to \mathcal{B}_h}} \qquad \qquad q\in \left\{ +,-\right\}
\end{align}
where $\mathcal{A}_{h_q}$ and $ \mathcal{B}_h$ are  given by 
\begin{equation*}
\mathcal{A}_{h_q}=- \frac{1}{2}q \epsilon_{h_q}k^2_{e\left(h\right)}\frac{\partial}{\partial k_{e\left(h_q\right)\mu_{e\left(h\right)}}} \qquad \qquad  \mathcal{B}_h=\eta^{\mu_{e\left(h_+\right)}\mu_{e\left(h_-\right)}}
\end{equation*}

 The striking feature of equation \eqref{PhitoYM} is that the  integrand $I_{\Upsilon_{l,n}}$ in \eqref{PhitoYM} can be replaced by the renormalised integrand  $I^R_{\Upsilon_{l,n}}$. The replacement produces  the renormalised integrand  ${\tilde I}_{\Gamma_{l,n}}^R$ of the quantum gauge theory. Thus we get the renormalised quantum gauge theory from the renormalised cubic quantum field theory.  This implies that a covariant quantization of gauge fields is possible without the need of introducing ghost fields.  
 
 
 
\section{Gauge theory amplitudes as sum over trivalent diagrams}\label{offamplitudes}

The relation between gauge theory amplitudes and the trivalent diagrams in a scalar quantum field theory has close resemblance with the relation between gauge theory and Witten's cubic string field theory. Although the Feynman diagrams in cubic string field theory are all trivalent, it correctly reproduces the scattering amplitudes in gauge theory in the field theory limit \cite{Taylor:2002bq}.  This is equivalent to the statement that the  low  energy classical effective action can be identified as the leading terms in the $\alpha'\to 0$ of the string field theory Wilsonian effective action  \cite{Sen:2016qap,Coletti:2003ai,Berkovits:2003ny}. The availability of the low energy effective action for massless fields of open string theory enables us to extract the low energy physics without doing direct string theory calculations which are usually challenging. However, we will argue that  Witten's cubic string field theory is capable of  unraveling the rich mathematical structures hidden in the gauge theory amplitudes.

\subsection{Witten's open string field theory}

Consider   $N$ number of coincident $Dp$ branes in a 26-dimensional flat spacetime. This system can be studied using Witten's cubic  string field theory \cite{Witten:1985cc}. The elementary degrees of freedom of string field theory are the string fields. A string field is an arbitrary state in the Fock space $\mathcal{H}$ of the world sheet conformal field theory. The world sheet conformal field theory describing  bosonic open string in a flat $26$-dimensional background is given by a sigma model of $26$ free scalars $X^{\mu}$ together with a $bc$ ghost system of central charge $-26$ with world-sheet action  
\begin{equation}\label{waction}
\frac{1}{2\pi \alpha'}\int d^2z \partial X^{\mu}\bar \partial X_{\mu} +\frac{1}{2\pi}\int d^2 z \left(b\bar \partial c +\bar b \partial \bar c \right).
\end{equation}
wher $\alpha'$ is the inverse of the string tension. The world-sheet BRST charge is given by 
\begin{equation}\label{BRSTCharge}
Q_B=\frac{1}{2\pi\text{i}}\oint \left( dz j_B-d\bar z \tilde j_B \right)
\end{equation}
where $j_B$ is the holomorphic BRST current 
\begin{equation}\label{BRSTc}
j_B=cT^X+\frac{1}{2} cT^g+\frac{3}{2}\partial^2 c
\end{equation}
and $\tilde j_B$ is the anti-holomorphic BRST current.  

The Hilbert space $\mathcal{H}$ of a single-string of momentum $k$ is constructed by the action of the raising and lowering oscillators on the highest weight vector $|k\rangle$ defined by the annihilation conditions 
\begin{align}\label{oscillatorsann}
a_{n\geq 1}^{\mu}|k\rangle=0\qquad p^{\mu}|k\rangle=k^{\mu}|k\rangle \qquad b_{n\geq 0}^{\mu}|k\rangle=0\qquad c_{n\geq 1}^{\mu}|k\rangle=0 
\end{align}
and commutation relations
\begin{align}\label{oscillatorscomm}
 \left[a_{m}^{\mu}, a_{-n}^{\nu} \right]=\eta^{\mu\nu}\delta_{m,n} \qquad\qquad \left\{b_{m}, c_{-n} \right\}=\delta_{m,n}
\end{align}
 The Hermitian conjugates of these oscillators are 
\begin{equation*}
{a_n^{\mu}}^{\dagger}=a_{-n}^{\mu} \qquad {b_n}^{\dagger}=b_{-n} \qquad {c_n}^{\dagger}=c_{-n}
\end{equation*}
The vacuum $|0\rangle$ can be expressed in terms of the $SL\left(2,\mathbb{R} \right)$ invariant vacuum $|1\rangle$ as $|0\rangle=c_1|1\rangle$. The state $|k\rangle$ satisfy the following normalisation condition
\begin{equation}\label{innerp}
\langle k|c_0|p\rangle=\left(2\pi\right)^{26}\delta\left(k+p\right).
\end{equation}
 Then the Fock space $\mathcal{H}$ a single string is spanned by set of all $|\chi\rangle$ having the form
\begin{equation*}
|\chi\rangle=\cdots a_{n_2}a_{n_1}\cdots b_{k_2} b_{k_1}\cdots c_{l_2}c_{l_1}|k\rangle
\end{equation*}
where $n_i,k_i<0$ and $l_i\leq 0$. 

A classical string field having ghost number 1 can be written as linear combination of such states having equal number of $b$ and $c$ oscillators
\begin{equation}\label{classicalstringfield}
|\Phi\rangle =\int dk^{p+1} \left(\phi^a(k)+A^a_{\mu}(k)a_1^{\mu}+\Phi^a_{\beta}(k)a_1^{\beta}-\text{i}\alpha^a(k)b_{-1}c_0+\cdots \right)|k, a\rangle
\end{equation} 
where the index $\mu$ is from $0$ to $p$ and $\beta$ is from $p+1$ to $25$.  The state  $|k, a\rangle$ is given by
\begin{equation}\label{chanpaton}
|k, a\rangle=\sum\limits_{i,j=1}^N|k, ij\rangle\lambda_{ij}^a
\end{equation}
 The spacetime fields transforms as the adjoint under the $U(N)$ symmetry due to the following transformation property of the Chan-Paton factors  under the $U(N)$ transformation 
 \begin{equation*}
 \lambda^a\to U \lambda^aU^{-1} \qquad a=1,\cdots N^2-1
 \end{equation*}
  This global symmetry of the world-sheet theory is elevated to $U(N)$ gauge symmetry in spacetime. As a result, the fields $A^a_{\mu}(k)$ and $\Phi^{a}_{\beta}$ forms massless $(p+1)$-dimesnional gauge fields and massless scalar fields respectively in the adjoint representation of $U(N)$.  
  
     The gauge invariant classical action for Witten's string field theory can be constructed in terms of the string field $|\Phi\rangle$ \cite{Witten:1985cc}
\begin{align}\label{WOSFTactionO}
S=-\frac{1}{2}\left\langle V_2| \Phi, Q_B\Phi\right\rangle +\frac{g}{3}\langle V_3 |\Phi, \Phi, \Phi\rangle
\end{align}
where $|V_2\rangle, |V_3\rangle$ are states in $\mathcal{H}^2, \mathcal{H}^3$.   The  kinetic and the cubic interaction term can be understood as the inner product between states in  $\mathcal{H}^2$ and $\mathcal{H}^3$ respectively.  The explicit expressions of $|V_2\rangle$ and $|V_3\rangle$ are given in the following subsection.
 
 \subsection{Off-shell amplitudes in string field theory}

The  off-shell amplitudes in  open string theory can be computed by choosing a gauge and by evaluating the Feynman diagrams.  In the Feynman-Siegel gauge 
\begin{equation}\label{FSgauge}
 b_0|\Phi\rangle = 0
\end{equation}
the propagator for string fields satisfying the takes the form
\begin{equation}\label{stringpropagator}
|P\rangle= \frac{b_0}{L_0}| V_2\rangle
\end{equation}

  An arbitrary Feynman diagram $\Upsilon$  of Witten's string field theory in Feynman-Siegel gauge can be evaluated by following the algorithm presented in \cite{Taylor:2002bq}.  
  \begin{itemize}
\item Prepare the state $\langle V_{\Upsilon}|$ in the Hilbert space obtained by taking the $3\left|\Upsilon^{[0]}\right|$-fold  tensor product of the dual of $\mathcal{H}$
 \begin{align}\label{V3product}
 \langle V_{\Upsilon}|= \otimes_{v\in \Upsilon^{[0]}} \langle V_{h_-hh_+}|
 \end{align}
 where $\langle V_{h_-hh_+}|$ is the Witten vertex expressed using the notion of half-edge $h$, the predecessor half-edge $h_-$ and the successor half-edge $h_+$
 \end{itemize}
 \begin{align}\label{Vh3}
&\langle V_{h_-hh_+}|=\mathcal{N}\int \prod\limits_{i\in\left\{ h_-,h,h_+\right\}}\left(d^{p+1}k_i \langle k|^{(i)}c_0^{(i)}\right)\delta\left( \sum\limits_{j\in\left\{ h_-,h,h_+\right\}} \sqrt{\alpha'} k_j\right)\text{exp}\left(-\sum\limits_{\substack{m\geq 0\\n>0\\r,s\in\left\{ h_-,h,h_+\right\}}}b_{m}^{(r)}X^{rs}_{mn}c^{(s)}_{n}\right)\nonumber\\
&\quad\times \text{exp}\left(- \frac{1}{2}\sum\limits_{\substack{m>0\\n>0\\r,s\in\left\{ h_-,h,h_+\right\} }}a_{\mu,m}^{(r)}V^{rs}_{mn}a_{\mu,n}^{(s)}-\sum\limits_{\substack{n>0\\r,s\in\left\{ h_-,h,h_+\right\}}}\sqrt{\alpha'}k_{\mu}^{(r)}V^{rs}_{0n}a_{\mu,n}^{(s)}-\frac{1}{2}\sum\limits_{\substack{r\in\left\{ h_-,h,h_+\right\}}}\alpha'k_{\mu}^{(r)}V^{rr}_{00}k_{\mu}^{(r)}\right)
\end{align}   
where $\mathcal{N}$ is a normalization constant. The values of the Neumann coefficients $V_{mn}^{r,s}$ can be found in \cite{Gross:1986ia, Gross:1986fk,Cremmer:1986if, Ohta:1986wn}. The  BPZ conjugation matrix  $C_{mn}$ is given by
 \begin{equation}\label{cnm}
C_{nm}=\delta_{nm}\left(-1\right)^n.
\end{equation}

 \begin{itemize}
\item  Prepare the state $ | P_{\Upsilon}\rangle$ that belongs to the  $2\left|\Upsilon^{[1]}\right|$-fold  tensor product of $\mathcal{H}$ 
  \begin{align}\label{V2product}
 | P_{\Upsilon}\rangle= \otimes_{e\in \Upsilon^{[1]}} | V_{e\left(h\right) e\left(\tilde h\right)}\rangle
 \end{align}
 where the state $ | V_{e\left(h\right) e\left(\tilde h\right)}\rangle$ implements the BPZ-inner product of the states associated with the two half-edges $e\left(h\right)$ and $ e\left(\tilde h\right)$ of the edge $e$. It is given by
 
  \end{itemize}

 \begin{align}\label{V2e}
 | V_{e\left(h\right) e\left(\tilde h\right)}\rangle=\int d^{p+1}k~ \text{exp}&\left(\sum\limits_{\substack{m>0\\n>0}}\left( a_{\mu,m}^{(h)}C_{mn}a_{\mu,n}^{(\tilde h)}-b_{m}^{(h)}C_{mn}c^{(\tilde h)}_{n}-b_{m}^{(
 \tilde h)}C_{mn}c^{(h)}_{n}\right)\right)  \nonumber\\
 &\times \left( c_0^{(h)}+c_0^{(\tilde h)}\right)  | k\rangle_h \otimes | -k \rangle_{h}.
\end{align} 

   \begin{itemize}
 \item The final ingredient is the $\left|\Upsilon^{[1]}_{\text{ext}} \right|$-fold tensor product of the dual of $\mathcal{H}$ 
 \begin{align}\label{extstate}
 \langle E_{\Upsilon}|=\otimes_{e\in \Upsilon^{[1]}_{\text{ext}}} ~_{\tilde h}\langle S_{e\left(\tilde h \right)}|
 \end{align}
 where $\langle S_{e\left(\tilde h \right)}|$ is the external state inserted at the external leg $e$ of $\Upsilon$. 
 
 \item Using these ingredients express the integrand of the contribution to the off-shell amplitude from Feynman diagram $\Upsilon$ having $n$ external legs and $l$ loops as 
 \begin{align}\label{offshellG}
 \mathcal{S}_{ l,n}\left(| E_{\Upsilon}\rangle \right) = \langle V_{\Upsilon}|\otimes  \langle E_{\Upsilon}| \int_{0}^{\infty} \prod_{e\in\Upsilon^{[1]}}dT_e~e^{-\frac{1}{2} T_e \left(L_0^{\left(e\left(\tilde h\right)\right)}+L_0^{\left(e\left(h\right)\right)}\right)} b_0^{(h)} | P_{\Upsilon}\rangle
 \end{align}
 where $k_{C_i}^{\mu}$ is the loop  momentum associated with $i^{\text{th}}$ loop $C_i$.
 \end{itemize}

\subsection{Gauge theory  Feynman diagrams from string field theory} \label{algorithm}
 
 Assume that all the external states in the scattering process are the massless gauge fields in string field theory.  The field theory limit of this amplitude can be obtained by computing the Feynman diagrams in the low energy Wilsonian effective action describing the dynamics of $Dp$ brane given by
\begin{equation}\label{effectiveS}
S_{eff}=-\frac{1}{4g^2_{YM}}\int d^{p+1}x ~\text{Tr} \left( {\hat F}^{\mu \nu} {\hat F}_{\mu \nu} +2 \left(D_{\beta}\Phi^a\right)^2-\left[ \Phi^a,\Phi^b\right]^2 \right)
\end{equation}
It is possible to recover the usual Feynamn diagram expansion of the Yang-Mills amplitudes by following  prescription given below \cite{Sen:2016qap} without relying on the low energy Wislonian effective action.
 \begin{itemize}
 \item Divide the string propagator $|P\rangle$ into two parts $$\Pi |P\rangle+\hat\Pi |P\rangle$$ where $\Pi |P\rangle$ allows the propagation of only massless fields and $\hat\Pi|P\rangle$ allows only the propagation of all massive fields.
 \item  Declare the sum of all the trivalent diagrams having $m$ number of massless external states and $g$ loops with all the cubic Witten vertices glued using the propagator  $\hat\Pi|P\rangle$ as the contact interaction of $m$ massless fields having $g$ loops.
 \item Construct the Wilsonian effective theory Feynman diagrams by gluing the contact interactions thus obtained using the  $\Pi |P\rangle$, the string propgator projected to the massless sector. 
  \item Take the field theory limit  by considering only the leading terms in the $\alpha'$ expansion of the resulting amplitude to obtain the gauge theory amplitude.
 \end{itemize}
This implies that the integrand of the $l$ loop contribution to a gauge theory off-shell amplitude with $n$ external legs as a sum over all trivalent diagram having $n$ external legs and $l$ loops 
\begin{align}\label{Fexpansion}
\mathcal{F}_{l,n}\left(\xi_1^{\mu_1},\cdots,\xi_l^{\mu_l} \right)=\sum\limits_{\Upsilon{n,l}} \lim_{\alpha'\to 0}\mathcal{S}_{{l,n}}\left(\xi_1^{\mu_1},\cdots,\xi_l^{\mu_l} \right)
\end{align}
 where ${\mathcal{S}}_{l,n}$ is the integrand associated with the cubic string field theory diagram $\Upsilon_{l,n}$ with the external states chosen to be the same as that of the gauge theory diagrams.

\section{Corolla differential from string field theory}\label{CDSFT}

Let us evaluate the field theory limit of ${\mathcal{S}}_{l,n}$, the integrand associated with the cubic string field theory diagram $\Upsilon_{l,n}$.  The massless fields in the presence of $N$ coincident $Dp$ brane are $p+1$-dimensional gauge fields and  $25-p$ scalar fields in the adjoint reprentation $U(N)$. However, we will restrict our analysis to pure Yang-Mills by allowing  only  the massless $(p+1)$-dimesnional  gauge fields to propgaate along the propagators of the Feynamnn  diagrams and by discarding the diagrams describing their interation with the massless scalars.

\subsection{Emergence of Corolla differentials}

The simplest possible diagram is the cubic diagram $\Upsilon_{0,3}$  describing the contact interaction of three gauge fields $A^a_{\mu_{e(h_-)}}$, $A^b_{\mu_{e(h)}}$, $A^c_{\mu_{e(h_+)}}$. The corresponding contribution $\mathcal{S}_{0,3}$  is computed in appendix \ref{A3}. The $\alpha'\to 0$ limit of $\mathcal{S}_{0,3}$ is given by
\begin{align}\label{V3ED00}
 &\mathcal{F}_{0,3} =\sqrt{\alpha'}g\sum\limits_{r \in\left\{ h_-,h,h_+\right\}}\left(\eta\wedge\eta\right)^{\mu\mu_{e(r_-)}\mu_{e(r)}\mu_{e(r_+)}} {k}_{e(r)\mu} 
\end{align}
where  
\begin{equation*}
\left(\eta \wedge \eta\right)^{\alpha_1\alpha_2\alpha_3\alpha_4}= \eta^{\alpha_1\alpha_2}\eta^{\alpha_3\alpha_4}-\eta^{\alpha_1\alpha_4}\eta^{\alpha_2\alpha_3}
\end{equation*}
 In the left hand side of the equation, we suppressed the tensor indices and in the right hand side we suppressed the color factor and the  momentum delta function associated with the diagram  to avoid cluttering.  Interestingy, the cubic interaction of gauge fields can also be expressed as the action of the Corolla differential  associated with the cubic graph $\Upsilon_{0,3}$
\begin{align}\label{corollaemerges}
\mathcal{F}_{ {0,3}} =- {k}_{e\left( h_-\right)}^2{k}_{e\left( h\right)}^2 {k}_{e\left( h_+\right)}^2 C_{v}^{\mu_{e\left( h_-\right)}\mu_{e\left( h\right)}\mu_{e\left( h_+\right)}}\left({k}_{e\left( h_-\right)}^2{k}_{e\left( h\right)}^2 {k}_{e\left( h_+\right)}^2\right)^{-1}
\end{align}
where $C_{v}^{\mu_{e\left( h_-\right)}\mu_{e\left( h\right)}\mu_{e\left( h_+\right)}}$ is the Corolla differential operator given by 
\begin{align}\label{corolla1}
&C_{v}^{\mu_{e\left( h_-\right)}\mu_{e\left( h\right)}\mu_{e\left( h_+\right)}} =-\frac{\sqrt{\alpha'}g }{2} \sum\limits_{r \in\left\{ h_-,h,h_+\right\}}{k_{e\left(r\right)}}^2\left(\eta\wedge\eta\right)^{\mu\mu_{e(r_-)}\mu_{e(r)}\mu_{e(r_+)}}\frac{\partial}{\partial {k}_{e(r)\mu}}
\end{align}

A more non-trivial emergence of Corolla polynomial from string field theory can be seen in the case of tree level quartic diagrams. This requires evaulating a quartic tree level string field theory  diagram $\Upsilon_{0,4}$ with massive states propagating through the internal edge.  In principle, it can be approximately calculated by performing the summation over all the massive states in the internal egde with the help of numerics \cite{Coletti:2003ai}.  Since we are only interested in the leading $\alpha'$ terms in the quartic interaction, we can  implement this algorithm using the analytic method discussed in \cite{Berkovits:2003ny}.  The field theory limit of the quartic interaction of four gauge fields is given by
 \begin{align}\label{smnu}
 k_{e(h)}^{-2}D^{\mu_{e(h_-)}\mu_{e(h)}\mu_{e(h_+)}}_h\left( k_{e\left(h\right)}^2D^{\mu_{e(\tilde h_-)}\mu_{e(h)}\mu_{e(\tilde h_+)}}_h k_{e(h)}^{-2}\right)
\end{align}
The detailed computation can be found in appendix \ref{A4}.

  Consider a tree diagram $\Upsilon_{0,4}$ in  string field theory with 4 external gaige fields. Using the cubic and quartic interaction we can evaluate this diagram
\begin{align}\label{schannel}
\mathcal{S}_{0,4}\left(| E_{ \Upsilon_{0,4}}\rangle \right) =&\langle V_{ \Upsilon_{0,4}}|\otimes  \langle E_{ \Upsilon_{0,4}}| \frac{b_0^{(h(e_5))}}{L_0^{\left(h(e_5)\right)}} | P_{ \Upsilon_{0,4}}\rangle\nonumber\\
=&\langle E_{ \Upsilon_{0,4}}|\langle V_{ \Upsilon_{0,4}}|\Pi_{(h(e_5))} \frac{b_0^{(h(e_5))}}{L_0^{\left(h(e_5)\right)}} | P_{ \Upsilon_{0,4}}\rangle+  \langle E_{ \Upsilon_{0,4}}|  \langle V_{ \Upsilon_{0,4}}| \hat {\Pi}_{(h(e_5))} \frac{b_0^{(h(e_5))}}{L_0^{\left(h(e_5)\right)}} | P_{ \Upsilon_{0,4}}\rangle
\end{align} 
where  $| E_{\Upsilon_{0,4}}\rangle=\otimes_{e\in \Upsilon^{[1]}_{\text{ext}}} ~a^{(h)}_{\mu_{e(h)}-1}| k_{e\left( h \right)}\rangle_{ h} $.   In the second line we divided the contribtuion from  $\Upsilon_{0,4}$ into two pieces, one with only the massless states propgating along the internal edge and the other with only the massive states propgating along it.  In the $\alpha'\to 0$ limit, the first term becomes the quartic tree diagram with an internal edge in pure Yang-Mills, and the second term mathces with \eqref{smnu}, the one by third of the Yang-Mills quartic interaction.

 Let us study the field theory limit of \eqref{schannel}.   Interestingly, the Corolla polynomial associated with $\Upsilon_{0,3}$ can be decomposed as follows
\begin{align}\label{fcubic}
 {k}_{e\left( h_-\right)}^2{k}_{e\left( h\right)}^2 {k}_{e\left( h_+\right)}^2 &C_{v}^{\mu_{e\left( h_-\right)}\mu_{e\left( h\right)}\mu_{e\left( h_+\right)}}\left({k}_{e\left( h_-\right)}^2{k}_{e\left( h\right)}^2 {k}_{e\left( h_+\right)}^2\right)^{-1} = \sum\limits_{r \in\left\{ h_-,h,h_+\right\}}k_{e\left(r\right)}^2 D^{\mu_{e(r_-)}\mu_{e(r)}\mu_{e(r_+)}}_rk_{e\left(r\right)}^{-2}
\end{align}
where 
\begin{equation*}
D^{\mu_{e(r_-)}\mu_{e(r)}\mu_{e(r_+)}}_r=- \frac{\sqrt{\alpha'}g }{2}k_{e\left(r\right)}^2\left(\eta\wedge\eta\right)^{\mu\mu_{e(r_-)}\mu_{e(r)}\mu_{e(r_+)}}\frac{\partial}{\partial {k}_{e(r)\mu}}
\end{equation*}
This decomposition reaveal the emergence of a surprising mathematical structure of  string field theory diagrams  in the field theory limit 
 \begin{align}\label{Fschannel}
\mathcal{F}_{ {0,4}}=& \sum\limits_{\substack{r \in\left\{ h_-,h, h_+\right\}\\ s \in\left\{ \tilde h_-,\tilde h,\tilde h_+\right\}}}k_{e\left(h\right)}^{-2} \left( k_{e\left(r\right)}^{2}D^{\mu_{e(r_-)}\mu_{e(r)}\mu_{e(r_+)}}_r k_{e\left(r\right)}^{-2}\right)\left(k_{e\left(s\right)}^{2}D^{\mu_{e(s_-)}\mu_{e(s)}\mu_{e(s_+)}}_s k_{e\left(s\right)}^{-2}\right) \nonumber\\
&\qquad+ k_{e(h)}^{-2}D^{\mu_{e(h_-)}\mu_{e(h)}\mu_{e(h_+)}}_h\left( k_{e\left(h\right)}^2D^{\mu_{e(\tilde h_-)}\mu_{e(h)}\mu_{e(\tilde h_+)}}_h k_{e(h)}^{-2}\right) \nonumber\\
=& \sum\limits_{\substack{r \in\left\{ h_-,h,h_+\right\}\\ s \in\left\{ \tilde h_-,h,\tilde h_+\right\}}} k_{e\left(h\right)}^{-2} \left( k_{e\left(r\right)}^{2}D^{\mu_{e(r_-)}\mu_{e(r)}\mu_{e(r_+)}}_r k_{e\left(s\right)}^{2}D^{\mu_{e(s_-)}\mu_{e(s)}\mu_{e(s_+)}}_s\right)k_{e\left(r\right)}^{-2}k_{e\left(s\right)}^{-2} \nonumber\\
= &\prod_{\tilde e\in \Upsilon_{0,4,\text{ext}}^{[1]}}k_{\tilde e}^{2} \left( \prod_{v\in \Upsilon_{0,4}^{[0]}}C^{\mu_{e(h_-)}\mu_{e(h)}\mu_{e(h_+)}}_{v}\right)\prod_{e\in \Upsilon_{0,4}^{[1]}}k_{e}^{-2}
\end{align} 
This implies that the $\alpha'\to 0$ limit of Witten's open string field theory  diagram $\Upsilon_{0,4}$ can be obtained by the action of the Corolla differential associated with $\Upsilon_{0,4}$  on the integrand of the same diagram in a scalar quantum field theory.

 \subsection{Corolla differentials for tree diagrams}
 
The relaion between Corolla differential and string field theory diagrams can be generalized to arbitrary tree diagrams. Assume that $\mathcal{S}_{0,n}$ is the contribtion to off-shell amplitude from an arbitrary tree diagram $\Upsilon_{0,n}$ with $n$ external massless gauge fields
 \begin{align}\label{TreeoffshellG}
 \mathcal{S}_{0,n}\left(| E_{ \Upsilon_{0,n}}\rangle \right) =\langle V_{\Upsilon_{0,n}}|\otimes  \langle E_{\Upsilon_{0,n}}|  \prod_{e\in\Upsilon_{0,n,\text{int}}^{[1]}}\frac{b_0^{(h(e))} }{L_0^{(h(e))}}| P_{\Upsilon_{0,n}}\rangle
 \end{align}
 We can divide $ \mathcal{S}_{0,n}$ into several pieces by distingushing whether massless states or massive states are flowing along the internal edges of $\Upsilon_{0,n}$.
  \begin{align}\label{SplitTreeoffshellG}
 \mathcal{S}_{0,n}\left(| E_{ \Upsilon_{0,n}}\rangle \right) &=\sum_{\mathcal{R}\in \mathcal{T}^{[1]}_{\Upsilon_{0,n}}}\langle V_{\Upsilon_{0,n}}|\otimes  \langle E_{\Upsilon_{0,n}}| \left( \prod_{\tilde e\in \mathcal{R}^c}  \Pi_{\tilde e}\prod_{\hat e\in \mathcal{R}} \hat \Pi_{\hat e}\right) \prod_{e\in\Upsilon_{0,n,\text{int}}^{[1]}}  \frac{b_0^{(h(e))} }{L_0^{(h(e))}}| P_{\Upsilon_{0,n}}\rangle
 \end{align}
where $\mathcal{T}^{[1]}_{\Upsilon_{0,n}}$ is the set of all the subsets of internal edges in  $\Upsilon_{0,n}$.  The set $\mathcal{R}^c$ is the complement of $\mathcal{R}$ in the set $\Upsilon_{0,n,\text{int}}^{[1]}$ of all the internal edges  of  $\Upsilon_{0,n}$
\begin{equation*}
\mathcal{R}\cup \mathcal{R}^c=\Upsilon_{0,n,\text{int}}^{[1]}
\end{equation*}
 In the field theory limit, the leading interaction terms in the Wilsonian effective action for the massless fields are  cubic and quartic interactions. This implies that in this limit, the leading part of  $\mathcal{S}_{0,n}$ is given by an expression that matches with \eqref{SplitTreeoffshellG}  except that in the summation the set $\mathcal{T}^{[1]}_{\Upsilon_{0,n}}$ is replaced by another set $\mathcal{U}^{[1]}_{\Upsilon_{0,n}}$. It is defined as the set of all the subsets of internal edges in  $\Upsilon_{0,n}$ having no common incident vertex. We can associate a diagram $\Upsilon_{\mathcal{X}}$ with each element $\mathcal{X}$ in the set $\mathcal{U}^{[1]}_{\Upsilon_{0,n}}$. The diagram $\Upsilon_{\mathcal{X}}$ has the property that along the edges in $\mathcal{X}$ only the massive states propgates. We shall call these edges as the marked edges\footnote{This can be understood as the string fielld theory interpretation of the marked edges discussed in \cite{Kreimer:2012jw}}. 
  \begin{align}\label{FieldSplitTreeoffshellG}
\mathcal{F}_{0,n}&= \lim_{\alpha'\to 0} \mathcal{S}_{0,n}\left(| E_{ \Upsilon_{0,n}}\rangle \right)\nonumber\\
&=\sum_{\mathcal{R}\in \mathcal{U}^{[1]}_{\Upsilon_{0,n}}}\lim_{\alpha'\to 0} \langle V_{\Upsilon_{0,n}}|\otimes  \langle E_{\Upsilon_{0,n}}| \left( \prod_{\tilde e\in \mathcal{R}^c}  \Pi_{\tilde e}\prod_{\hat e\in \mathcal{R}} \hat \Pi_{\hat e}\right) \prod_{e\in\Upsilon_{0,n,\text{int}}^{[1]}}  \frac{b_0^{(h(e))} }{L_0^{(h(e))}}| P_{\Upsilon_{0,n}}\rangle
 \end{align}
 
We claim that the relaion between Corolla differential and string field theory diagram  can be generalised to arbitrary tree diagrams in the following way
 \begin{align}\label{FGamama}
\mathcal{F}_{0,n}= \delta\left(k_{\Upsilon_{0,n}}\right)\prod_{\tilde e\in \Upsilon_{0,n,\text{ext}}^{[1]}}k_{\tilde e}^{2} \left( \prod_{v\in \Upsilon_{0,n}^{[0]}}C^{\mu_{e(h_-)}\mu_{e(h)}\mu_{e(h_+)}}_{v}\right)\prod_{e\in \Upsilon_{0,n}^{[1]}}k_{e}^{-2}
\end{align} 
where $\delta\left(k_{\Upsilon_{0,n}}\right)$ is the momentum conservation delta function for the graph $\Upsilon_{0,n}$. We will prove this claim using mathematical  induction. This requires proving this claim for $n=4$ which is already done in the previous section. Therefore, we are only left to prove this claim for general $n$ by assuming that \eqref{FGamama} is true for the graph $\Upsilon_{n-1}$. 

The graph  $\Upsilon_{0,n}$ can be constructed by attaching a trivalent  vertex to one of the external edge of the graph $\Upsilon_{0,n-1}$.  Denote the new internal edge and vertex introduced by this process as $e_n$ and $v_n$ respectively. This edge can be a maked edge or an unmarked edge. We will deote $\Upsilon_{0,n}$ as $\hat \Upsilon_{0,n}$   if the edge $e_n$ is a marked edge and $\tilde\Upsilon_{0,n}$ otherwise. The edge $e_n$ is assumed to connect vertex denoted as $v_{n-1}$ and $v_n$, and the edges incident on $v_{n_1}$ are $e_{n-1}\left(h\right)=e_n\left(h\right), e_{n-1}\left(h_-\right)$ and $e_{n-1}\left(h_+\right)$.  The graph obtained by removing the vertices $v_{n-1}$ and $v_{n-2}$ from $\Upsilon_{0,n}$ will be called as $\Upsilon_{0,n-2}$. Then we have
  \begin{align}\label{Recursion}
& \mathcal{F}_{0,n} = \lim_{\alpha'\to 0}\mathcal{S}_{0,n-1}\left(| E_{ \Upsilon_{0,n-1}}\rangle \right) \frac{1}{k^2_{e_n\left(h\right)}}\left(\sum\limits_{r \in\left\{ h_-,h,h_+\right\}}k_{e_n\left(r\right)}^2 D^{\mu_{e_n(r_-)}\mu_{e_n(r)}\mu_{e_n(r_+)}}_rk_{e_n\left(r\right)}^{-2}\right) \nonumber\\
 &+ \lim_{\alpha'\to 0}\mathcal{S}_{0,n-2} k^{-2}_{e_{n-1}\left(h_-\right)}\left( k_{e_n(h)}^{-2}D^{\mu_{e_{n-1}(h_-)}\mu_{e_n(h)}\mu_{e_{n-1}(h_+)}}_h\left( k_{e_n\left(h\right)}^2D^{\mu_{e_n( h_-)}\mu_{e_n(h)}\mu_{e_n( h_+)}}_h k_{e_n(h)}^{-2}\right)\right) \nonumber\\
 &=\left(\prod_{\tilde e\in \Upsilon_{0,n,\text{ext}}^{[1]}}k_{\tilde e}^{2} \left( \prod_{v\in \Upsilon_{0,n}^{[0]}}\left(\sum_{r \in\left\{ h_-,h,h_+\right\}} D^{\mu_{e(r_-)}\mu_{e(r)}\mu_{e(r_+)}}_{v,e(r)}\right)\right)\prod_{e\in \Upsilon_{0,n}^{[1]}}k_{e}^{-2}\right)
 \end{align}
 Here we used the fact the in the $\alpha'\to 0$ limit the leading contribtions do not come from diagrams with more than one maked edges that are incident on the same vertex. This proves the claim \eqref{FGamama}. 
 
  \subsection{Corolla differentials for loop diagrams}

We can extend this analaysis to arbitrary loop diagrams. Feynman diagrams with loops, unlike the tree diagrams, can have internal lines that belong to the loops with ghost/ anti-ghost states  propagating  along them, even if the all the external states are physical components of the gauge fields.  The field theory limit of the cubic contact interaction of a gauge  field with ghost and anti-ghost fields  can be expressed as
\begin{align}\label{McubiAint}
 V_{Ag\bar g }    &=\sqrt{\alpha'}g k_{e\left(h_-\right)}^2k_{e\left(h\right)}^2k_{e\left(h_+\right)}^2 \mathcal{G}_v^{\mu_{e\left(h\right)}}  k_{e\left(h_-\right)}^{-2}k_{e\left(h\right)}^{-2}k_{e\left(h_+\right)}^{-2}
\end{align}
Details of the computation is given in appendix \ref{AGG}. The differential operator  $ \mathcal{G}_v^{\mu_{e\left(h\right)}}$ is given by
 \begin{align}
 \mathcal{G}_v^{\mu_{e\left(h\right)}}=\frac{1}{2}\left(k_{e\left(h_-\right)}^2\frac{\delta}{\delta k_{e\left(h_-\right)\mu_{e\left(h\right)}}}-k_{e\left(h_+\right)}^2\frac{\delta}{\delta{k}_{e\left(h_+\right)\mu_{e\left(h\right)}}}\right) 
 \end{align}

  Consider  Feynman diagram $\Upsilon_{1,n}$ in  Witten's open string field theory with $n$ external legs and a loop denoted as $C_1$. The edges that incident on a vertex that belongs to the loop $C_1$ are either an external leg of  $\Upsilon_{1,n}$ or part of a connected tree diagram denoted as $\Upsilon_{\not{C_1},n}$ obtained by removing the loop from it.  In the $\alpha'\to 0$ limit there are only quartic interactions involving four gauge fields and cubic interactions involving either three gauge fields or a gauge, ghost  and anti-ghost fields. As a result,  along all the edges of $\Upsilon_{\not{C_1},n}$ only gauge fields are the only massless states that can propagate. However, along the edges that belongs to  the loop $C_1$, all massless excitations, including ghost/anti-ghost fields,  can propagate. However, if ghost/anti-ghost field appear in one of the edge that belongs to the loop, then  all the remaing edges in the loop also allows only the propagation  of  only these states.  This implies that 
   \begin{align}\label{FieldoneLoopoffshellG}
 \lim_{\alpha'\to 0}\mathcal{S}_{1,n}\left(| E_{ \Upsilon_{1,n}}\rangle \right) &= H^0\left(| E_{ \Upsilon_{1,n}}\rangle \right)-H^1\left(| E_{ \Upsilon_{1,n}}\rangle \right)
 \end{align}
 where $H^0$ is the contribution when no ghost/anti-ghost appear in any of the internal edges of $\Upsilon_{1,n}$ and $H^1$ is the contribution when only the ghost /anti-ghost field propagate along the loop $C_1$. Clearly, when  the  massless states that  propagate along the diagram are only the gauge fields, all the vertices in the diagram can be replaced by the differentail  associated with the $A^3$ vertex,  and hence $H^0$ is given by
  \begin{align}\label{FLoopGamama}
H^0= \delta\left(k_{\Upsilon_{1,n}}\right)\prod_{\tilde e\in \Upsilon_{1,n,\text{ext}}^{[1]}}k_{\tilde e}^{2} \left( \prod_{v\in \Upsilon_{1,n}^{[0]}}C^{\mu_{e(h_-)}\mu_{e(h)}\mu_{e(h_+)}}_{v}\right)\prod_{e\in \Upsilon_{1,n}^{[1]}}k_{e}^{-2}
\end{align}  
where $k_{C_1}$ is the loop momentum associated with the loop $C_1$. However, when  only the ghost /anti-ghost field propagate along the loop $C_1$, the vertices that are  part of the loop must be replaced by the differentail  associated with the $Ag\bar g$ vertex. The cubic contact  $Ag\bar g $ SFT Feynmann diagram in the field theory limit can be replaced  by the action of a differential operator $\mathcal{G}_v$. Therefore, $H^1$ is given by
  \begin{align}\label{FH1}
H^1= \delta\left(k_{\Upsilon_{1,n}}\right)\prod_{\tilde e\in \Upsilon_{1,n,\text{ext}}^{[1]}}k_{\tilde e}^{2} \prod_{v\in \Upsilon_{\not{C_1},1,n}^{[0]}}C^{\mu_{e(h_-)}\mu_{e(h)}\mu_{e(h_+)}}_{v} \prod_{v\in C_1^{[0]}}\mathcal{G}_v^{\mu_{e(h)}}\prod_{e\in \Upsilon_{1,n}^{[1]}}k_{e}^{-2}
\end{align}



It is straightforward to  generalize the analysis in the previous subsection to  a trivalent Feynman graph $\Upsilon_{l,n}$ with $l$ loops and $n$ external edges.  Assume that these loops can be generated by a set of basis loops denoted by $\mathbb{C}\left(\Upsilon_{l,n}\right)$.  We denote the set of external edges as $\Upsilon_{l,n,\text{ext}}^{[1]}=\left\{E \right\}$, the set of vertices that are part of  a loop $C_j$ as $C^{[0]}_j$  and the internal edges that are not part of loops  $C_1,\cdots,C_m$ as $\Upsilon_{\not{C_1}\cdots \not{C_m}, l,n}^{[1]}$.  Then the leading terms in the $\alpha \to 0$ limit of a string field theory Feynman diagram $\Upsilon_{l,n}$ with all external states chosen to be gauage fields is given by
    \begin{align}\label{FLoopoffshellG}
 \lim_{\alpha'\to 0}\mathcal{S}_{l,n}\left(| E_{ \Upsilon_{l,n}}\rangle \right) &= \sum_{i}\left(-1\right)^iH^i\left(| E_{ \Upsilon_{l,n}}\rangle \right)
 \end{align}
 where $H_i$ is the contribution from $\Upsilon_{l,n}$ with gosts and anti-ghost fields propgating in $i$ number of loops $C_1,\cdots, C_i$ 
 \begin{align}\label{Hi}
 H_i=\prod_{\tilde e\in \Upsilon_{l,n,\text{ext}}^{[1]}}k_{\tilde e}^{2} \delta\left(k_{\Upsilon_{l,n}}\right)\prod_{v\in \Upsilon_{\not{C_1}\cdots \not{C_m},l,n}^{[0]}}C^{\mu_{e(h_-)}\mu_{e(h)}\mu_{e(h_+)}}_{v} \prod_{v\in C_1^{[0]},\cdots, C_i^{[0]}}\mathcal{G}_v^{\mu_{e(h)}}\prod_{e\in \Upsilon_{l,n}^{[1]}}k_{e}^{-2}
 \end{align}
 The factor $\left(-1\right)^i$ associated with $H_i$ is due to the fact that gost and antighost  are anti-commuting fields. Consequently, the field theory limit of a string field theory loop diagram $\Upsilon_{l,n}$ with gauge fields as external states is given by 
   \begin{align}\label{Final}
\mathcal{F}_n&=\sum_l\sum_{\Upsilon_{l,n}} \text{i}^{|\Upsilon_{l,n}^{[1]}|} \text{color}\left(\Upsilon_{l,n}\right)g_{\text{YM}}^{|\Upsilon_{l,n}^{[0]}|}\prod_{\tilde e\in \Upsilon_{l,n,\text{ext}}^{[1]}}k_{\tilde e}^{2}  \int \prod_{j=1}^m dk_{C_j}  \delta\left(k_{\Upsilon_{l,n}}\right) ~\mathcal{D}\left(\Upsilon_{l,n}\right)\prod_{e\in \Upsilon_{l,n}^{[1]}}k_{e}^{-2} \end{align}
 where  $g_{YM}=\frac{g}{\sqrt{2}}$ is the Yang-Mills coupling. The $\text{i}$ factors associated with the propgators and the color factor $\text{color}\left(\Upsilon_{l,n}\right)$ associated with the diagram $\Upsilon_{l,n}$ are explicitly written for complenetss. The differential $\mathcal{D}\left(\Upsilon_{l,n}\right)$ mathces precisely with the Corolla differential \eqref{CtoD}.

 \section{Conclusion}\label{conclusion}
 
In this paper, we showed that the result of Kreimer, Sars and van Suijlekom derived in \cite{Kreimer:2012jw} using graph homology, which states that the amplitudes in a generic non-abelian gauge theory can be constructed by transmuting the  integrands of trivalent scalar quantum field theory  diagrams using Corolla differentials, naturally emerges from Witten's open string field theory in the field theory limit. We demonstrated that the action of Corolla differential on a trivalent scalar quantum field theory diagram produces the field theory limit of the same Witten's open string field theory diagram with all external states chosen to be gauge fields. Since the Corolla differentials are intimately tied to the  graph homology, this result alludes to the astounding interplay between  graph homology and string field theory.  It will be interesting explore this profound link further by including the stringy corrections and study the possibilty of formulating a stringy generalization of graph homology. If such a generalization exists, then that may be useful in systematically deriving the Wilsonian effective action of string theory.

   \acknowledgments
   
We would like to thank Alok Laddha and Ashoke Sen for the insightful discussions and suggestions. We also thank Sujay Ashok, Rajesh Gopakumar, R. Loganayagam,  Suvrat Raju, Aninda Sinha, Nemani V. Suryanarayana and Spenta Wadia for helpful discussions. 

\appendix 

\section{Cubic interaction $A^3$}\label{A3}

An arbitrary state in the world-sheet Fock space can be obtained  from the coherent state  defined for a half-edge $h$
  \begin{align}\label{Gstateexp}
 |G_h\rangle&=\text{exp}\left(\sum\limits_{\mu, m>0} J_{e\left(h\right) \mu_{e\left(h\right)},m}a_{\mu_{e\left(h\right)},-m}^{h}-\sum\limits_{m\geq 0}\mathcal{J}_{e\left(h\right)  bm}b^{h}_{-m}+\sum\limits_{m>0}\mathcal{J}_{e\left(h\right)  cm}c^{h}_{-m}\right)|k_{e\left(h\right)}\rangle
  \end{align}
 For example, the state $\prod\limits_{m,n,q}a_{\mu m}b_{n}c_{q}|p\rangle$ obtained by acting $| G\rangle$ with the  differential operator 
\begin{equation}\label{diffoperator}
\prod_{m,n,q}\frac{\partial}{\partial J_{\mu m}}\frac{\partial}{\partial \mathcal{J}_{bn}}\frac{\partial }{\partial \mathcal{J}_{cq}}
\end{equation}
and setting $J,\mathcal{J}_b,$ and $\mathcal{J}_c$ to 0.  This imply that the cubic interaction 
\begin{equation*}
\langle  V^v_{h_-h h_+}| | G_{h_-}\rangle|  G_{h}\rangle|  G_{h_+}\rangle
\end{equation*}
 of any three states can be computed from the following cubic interaction of  the coherent states  $ |G_{h_-}\rangle,  |G_h\rangle$ and $  |G_{h_+}\rangle$ by appropriately differentiating  with respect to the variables $J,\mathcal{J}_b,$ and $\mathcal{J}_c$. This cubic interaction can be calculated using the identities for the inner product between two squeezed state of the bosonic $a$ oscillators \eqref{MidentityG1} and the fermionic $bc$-ghost system oscillators \eqref{Midentitybc}. It is given by
\begin{align}\label{V123R}
&\langle  V^v_{h_-h h_+}| | G_{h_-}\rangle|  G_{h}\rangle|  G_{h_+}\rangle=\mathcal{N} \delta\left( \sum\limits_{r\in \left\{h_-,h,h_+ \right\}} k_{e\left( r\right)}\right)\text{exp}\left(\frac{1}{2}\alpha'\sum\limits_{r\in \left\{h_-,h,h_+ \right\}}{{k}_{e\left(r\right)\mu_{e\left(r\right)}}{ V}^{rr}_{00}{{k}_{e\left(r\right)\mu_{e\left(r\right)}}}}\right)\nonumber\\
& \text{exp}\left( -\sum\limits_{\substack{r,s\in  \left\{h_-,h,h_+ \right\}\\m>0,n\geq 0}}{\mathcal{J}}_{e\left(r\right)c,m} {  X}^{rs}_{mn}{\mathcal{J}}_{e\left(s\right)b,n}+ \frac{1}{2}\sum\limits_{\substack{{r, s\in \left\{h_-,h,h_+ \right\}}\\ n,m>0}}{J}_{e\left(r\right)\mu_{e\left(r\right)},n}{  V}^{rs}_{nm}{J}_{e\left(s\right)\mu_{e\left(s\right)},m}\right)  \nonumber\\
& \text{exp}\left(-\sum\limits_{\substack{{r, s\in \left\{h_-,h,h_+ \right\}}\\ m>0}}{k}_{e\left(r\right)\mu_{e\left(r\right)}}{  V}^{rs}_{0m}{J}_{e\left(s\right)\mu_{e\left(s\right)},m}\right) 
\end{align}


The contribution from the Feynman diagram $\Upsilon_{0,3}$ describing the contact interaction of three gauge fields $A_{\mu_{e(h_-)}}$, $A_{\mu_{e(h)}}$, $A_{\mu_{e(h_+)}}$ in a Non-abelian Yang-Mills theory  can be obtained by by differentiating  $\langle  V^v_{h_-h h_+}| | G^1_{h_-}\rangle|  G^1_{h}\rangle|  G^1_{h_+}\rangle$  with respect to $J_{e\left(h_-\right)\mu_{e\left(h_-\right)},1}$, $J_{e\left(h\right)\mu_{e\left(h\right)},1}$, $J_{e\left(h_+\right)\mu_{e\left(h_+\right)},1}$ 
\begin{align}\label{cubiAint}
\mathcal{F}_{{0,3}} =g\lim_{\alpha'\to 0}\frac{\delta }{\delta J_{e\left(h_-\right)\mu_{e\left(h_-\right)},1} } \frac{\delta }{\delta J_{e\left( h\right)\mu_{e\left( h\right)},1} }\frac{\delta }{\delta J_{e\left(h_+\right)\mu_{e\left(h_+\right)},1} }\langle  V^v_{h_-h h_+}| | G^1_{h_-}\rangle|  G^1_{h}\rangle|  G^1_{h_+}\rangle \Bigg |_{J,\mathcal{J}_{b,c}=0}
\end{align}
where the coherent state $ |G^1_h\rangle$ is given by 
  \begin{align}\label{Gstateexp1}
 |G^1_h\rangle=\text{exp}\left( J_{e\left(h\right) \mu_{e\left(h\right)},1}a_{\mu_{e\left(h\right)},-1}^{h}\right)|k_{e\left(h\right)}\rangle
  \end{align}
  
  Suppose that $f(x)$ is a function given by
\begin{align}\label{xy} 
f(x,z)=A(x,z)\frac{\partial^n}{\partial y^n}e^{-\left( y+cx+z\right)^2}|_{y=0}
\end{align}
 By applying he chain rule we can see that  $f(x)$  can also be written as  
\begin{align}\label{xytoxw} 
f(x,z)=\frac{1}{c^n}A(x,z)\frac{\partial^n}{\partial x^n}e^{-\left(cx+z\right)^2}
\end{align}
In order to make use of this fact, let us  express $\langle  V^v_{h_-h h_+}| | G^1_{h_-}\rangle|  G^1_{h}\rangle|  G^1_{h_+}\rangle$ as  
  \begin{align}\label{V123R1}
&\langle  V^v_{h_-h h_+})| | G^1_{h_-}\rangle|  G^1_{h}\rangle|  G^1_{h_+}\rangle= \mathcal{N} \delta\left( \sum\limits_{r\in \left\{h_-,h,h_+ \right\}} k_{e\left( r\right)}\right) \text{exp}\left(\frac{1}{2}\alpha'{{k}_{e\left(r\right)\mu_{e\left(r\right)}}{ V}^{rr}_{00}{{k}_{e\left(r\right)\mu_{e\left(r\right)}}}}\right)  \nonumber\\
& \text{exp}\left( \frac{1}{2}\left({J}_{e\left(r\right)\mu_{e\left(r\right)},1}^T-\sqrt{\alpha'}{k}_{e\left(q\right)\mu_{e\left(q\right)},1}V_{01}^{qp}{V^{-1}_{11}}^{pr}\right){  V}^{rs}_{11}\left({J}_{e\left(s\right)\mu_{e\left(s\right)},1}-\sqrt{\alpha'}{V^{-1}_{11}}^{sl}V_{01}^{ld}{k}_{e\left(d\right)\mu_{e\left(d\right)},1}\right)\right) \nonumber\\
&\text{exp}\left(-\frac{1}{2}\alpha'\left({k}_{e\left(q\right)\mu_{e\left(q\right)},1}V_{01}^{qp}{V^{-1}_{11}}^{pl}V_{01}^{ld}{k}_{e\left(d\right)\mu_{e\left(d\right)},1}\right)\right)
\end{align}
where 
\begin{equation*}
\bold{k}_{e\left(r\right)\mu_{e\left(r\right)}e\left(r\right),1}= \left({V^{-1}_{11}}V_{01}\right)^{rs}{k}_{e\left(d\right)\mu_{e\left(s\right)},1}
\end{equation*}
The matrices $V_{11}^{rs}$ and $V_{01}^{rs}$ are 
\begin{align}\label{V11V01}
V_{11}&=\frac{1}{27}\begin{bmatrix}
         5  & -16& -16 \\
    -16      &  5& -16  \\
      -16     & -16 &  5
\end{bmatrix} \qquad V_{01}= \frac{2\sqrt{2}}{3\sqrt{3}}\begin{bmatrix}
         0  & -1& 1  \\
1      &  0 & -1  \\
      -1     & 1 &  0
\end{bmatrix}  
\end{align}

This expression of  $\langle  V^v_{h_-h h_+}| | G_{h_-}\rangle|  G_{h}\rangle|  G_{h_+}\rangle$ can be compared with $f(x,z)$ if  the variable $y$ is  identified with $J_{e\left(r\right)\mu_{e\left(r\right)},1}$ and the variable $x$ is mapped to $\sqrt{\alpha'}\bold{k}_{e\left(r\right)\mu,1}$.  This implies the derivatives with respect to the generating parameters $J_{e\left(r\right)\mu_{e\left(r\right)}e\left(r\right),1}$ can be exchanged with the momentum derivatives $\sqrt{\alpha'}\bold{k}_{e\left(r\right)\mu,1}$. 
\begin{align}\label{KcubiAint}
\mathcal{F}_{{0,3}} =g\lim_{\alpha'\to 0}{\alpha'}^{-\frac{3}{2}}\frac{\delta }{\delta \bold{k}_{e\left(h_-\right)\mu_{e\left(h_-\right)},1} } \frac{\delta }{\delta \bold{k}_{e\left( h\right)\mu_{e\left( h\right)},1} }\frac{\delta }{\delta \bold{k}_{e\left(h_+\right)\mu_{e\left(h_+\right)},1} }\langle  V^v_{h_-h h_+}| | G^1_{h_-}\rangle|  G^1_{h}\rangle|  G^1_{h_+}\rangle \Bigg |_{J,\mathcal{J}_{b,c}=0}
\end{align}
Using chain rule  we get that 
\begin{align}\label{deltak}
  \frac{\delta }{\delta \bold{k}_{e\left( h\right)\mu_{e\left( h\right)},1} }=\frac{27\sqrt{3}}{14\sqrt{2}}\left( \frac{\delta}{ \delta {k}_{e\left( h_-\right)\mu_{e\left( h\right)}} }-\frac{\delta}{ \delta {k}_{e\left( h_+\right)\mu_{e\left( h\right)}} }\right)
\end{align}
The result obtained by subtituiting \eqref{deltak} in \eqref{KcubiAint} is given by
\begin{align}\label{V3ED00}
 &\mathcal{F}_{0,3} =\sqrt{\alpha'}g \delta\left( \sum\limits_{r\in \left\{h_-,h,h_+ \right\}} k_{e\left( r\right)}\right)\sum\limits_{r \in\left\{ h_-,h,h_+\right\}}\left(\eta\wedge\eta\right)^{\mu\mu_{e(r_-)}\mu_{e(r)}\mu_{e(r_+)}} {k}_{e(r)\mu} \nonumber\\
\end{align}
where $\left(\eta \wedge \eta\right)^{\alpha_1\alpha_2\alpha_3\alpha_4}= \eta^{\alpha_1\alpha_2}\eta^{\alpha_3\alpha_4}-\eta^{\alpha_1\alpha_4}\eta^{\alpha_2\alpha_3}$. The normalization constant is chosen to be 
\begin{equation}
\mathcal{N}=\left(\frac{3\sqrt{3}}{2\sqrt{2}}\right)^3
\end{equation}

\section{Cubic interaction $Ag\bar g$}\label{AGG}
The field theory limit of the cubic contact interaction of a gauge  field with ghost and anti-ghost fields also can be computed using the method discussed in appendix \ref{A3}. It is given by
\begin{align}\label{cubiAint}
 V_{Ag\bar g } &=g\lim_{\alpha'\to 0} \frac{\delta }{\delta J_{e\left( h\right)\mu_{e\left( h\right)},1} }\langle  V^v_{h_-h h_+}| c^{h_-}_{-1}|k_{e\left(h_-\right)}\rangle |  G^1_{h}\rangle b^{h_+}_{-1} |k_{e\left(h_+\right)}\rangle \Bigg |_{J,\mathcal{J}_{b,c}=0}\nonumber\\
  &=\sqrt{\alpha'}g \mathcal{N}X_{11}^{12}V_{01}^{12} \delta\left( \sum\limits_{r\in \left\{h_-,h,h_+ \right\}} k_{e\left( r\right)}\right)\eta^{\mu_{e\left(h\right)\nu_{e\left(h\right)}}}\left({k}_{e\left(h_-\right)\nu_{e\left(h\right)}}-{k}_{e\left(h_+\right)\nu_{e\left(h\right)}}\right) \nonumber\\
    &=\frac{\sqrt{\alpha'}g}{2} \delta\left( k_{\Upsilon_{0,3}}\right)k_{e\left(h_-\right)}^2k_{e\left(h\right)}^2k_{e\left(h_+\right)}^2\left(k_{e\left(h_-\right)}^2\frac{\delta}{\delta k_{e\left(h_-\right)\mu_{e\left(h\right)}}}-k_{e\left(h_+\right)}^2\frac{\delta}{\delta{k}_{e\left(h_+\right)\mu_{e\left(h\right)}}}\right) k_{e\left(h_-\right)}^{-2}k_{e\left(h\right)}^{-2}k_{e\left(h_+\right)}^{-2}
\end{align}
 where 
\begin{equation*}
\tilde{\bold{k}}_{e\left(h\right)\mu_{e\left(h\right)}e\left(h\right),1}={k}_{e\left(q\right)\mu_{e\left(q\right)}}V_{01}^{q2}{V^{-1}_{11}}^{22}
\end{equation*}
Here we used the following relation between the ghost and matter Neumann coefficients
\begin{align}
X_{11}^{12}=\left(V_{01}^{12}\right)^2
\end{align}

\section{Quartic interactions $A^4$} \label{A4}

The low energy effective action for the string field that consists of  massless string excitations can be systematically constructed by spilt the string field  $|\Phi\rangle$ into two pieces
\begin{equation}\label{split}
|\Phi\rangle = |\Psi\rangle+|\Theta \rangle
\end{equation}
 where $|\Psi\rangle$ contain only the massless excitations and $|\Theta\rangle$ contains all the massive excitations. Then the cubic action \eqref{WOSFTactionOSplit} becomes
\begin{align}\label{WOSFTactionOSplit}
S=&-\frac{1}{2}\left\langle V_2| \Psi, Q_B\Psi\right\rangle +\frac{g}{3}\langle V_3 |\Psi, \Psi, \Psi\rangle \nonumber\\
&-\frac{1}{2}\left\langle V_2|\Theta, Q_B\Theta\right\rangle+g\langle V_3 |\Psi, \Psi, \Theta\rangle+g\langle V_3 |\Psi, \Theta, \Theta\rangle+\frac{g}{3}\langle V_3 |\Theta, \Theta, \Theta\rangle
\end{align}
The higher order contact interaction for the massless string field $|\Psi\rangle$ can be obtained by integrating out the massive string field $|\Theta\rangle$. Since we are integrating out the massive string field, we are allowed to redfine $|\Theta\rangle$ as below
\begin{equation}\label{shift}
 |\Theta \rangle  = |\tilde \Theta \rangle-\frac{b_0}{L_0} \langle \Psi| \langle\Psi ||V_3\rangle
\end{equation}
This field redefinition removes the terms linear in $|\Theta\rangle$ from the action and produces the following  term which is quartic in $|\Psi\rangle$
\begin{equation}\label{quartic}
S^{(4)}\left( \Psi\right) = \frac{1}{2}\langle V_2|\frac{b_0}{L_0} \langle \Psi| \langle\Psi ||V_3\rangle Q_B\frac{b_0}{L_0} \langle \Psi| \langle\Psi ||V_3\rangle
\end{equation}
Since the ghost number 1 masless zero momentum fields satisfying Siegel gauge  are BRST exact states, they cannot propogate along the internal propgator due to the presence of $Q_B$. Therefore the quartic interaction takes the following form 
\begin{align}\label{quartic1}
S^{(4)}\left( \Psi\right) &= \frac{1}{2}\langle V_2|\frac{b_0}{L_0}\hat\Pi\langle \Psi| \langle\Psi ||V_3\rangle \langle \Psi| \langle\Psi ||V_3\rangle 
\end{align}
 where $\hat\Pi$ is the projection operator assosicated with the massive states in the Hilbert space.  Since the  zero momentum gauge fields are BRST exact states,  the tree level amplitude of the four zero momentum gauge field interaction vanishes. 
 
 The inner product in \eqref{quartic} between the two states produced by the Witten vertex can be mapped to a correlation function of four vertex operators using the notion of star product and  wedge states \cite{Rastelli:2000iu}.  The star product between state two arbitrary  states $|\Phi_1\rangle$ and $|\Phi_2\rangle$ is defined as 
 \begin{equation}\label{star}
 |\Phi_1 \rangle \star |\Phi_2\rangle= \langle \Phi_1|\langle \Phi_2 ||V_3\rangle
 \end{equation}
It is obtained by inserting the vertex operators $\mathcal{V}_{\Phi_1}, \mathcal{V}_{\Phi_2}$ at the points $z=\frac{1}{\sqrt{3}}$ and $z=-\frac{1}{\sqrt{3}}$  on the upper half-plane followed by the conformal transformation $f^{2\pi}(z)$. The conformal transformation
 \begin{equation}\label{nmap}
w=f^{\frac{2}{n}}(z)=\left(\frac{1+\text{i}z}{1-\text{i}z}\right)^{\frac{2\pi}{n}}=\text{tan}\left(\frac{2}{n}\text{arctan}~z\right)
\end{equation}
 sends the unit upper half disk in the $z$ plane to a wedge of angle $\frac{2\pi}{n}$ in the $w$ plane \cite{Rastelli:2000iu}. 

Let $U_n$ be the representation of the conformal transformation $f^{\frac{2}{n}}(z)$  on the string Hilbert space
\begin{equation}\label{Un}
U_n=\text{exp}\left(2~\text{ln} \frac{n}{2}\left( -\frac{1}{2}L_0+\sum\limits_{k=1}^{\infty}\frac{\left(-1\right)^k}{\left(2k-1 \right)\left(2k+1 \right)}L_{2k}\right) \right)
\end{equation}
 Then by using $U_3$ we can express the inner product \eqref{quartic} as the correlation function
\begin{align}\label{U3def}
&\left \langle c{\partial X_{\mu}}{\left({\sqrt{3}}\right)}c{\partial X_{\nu}}{\left(-{\sqrt{3}}\right)} U_3\frac{b_0}{L_0} U^{\dagger}_3 {\partial X_{\sigma}}{\left(-\frac{1}{\sqrt{3}}\right)}c{\partial X_{\sigma}}{\left(\frac{1}{\sqrt{3}}\right)}\right\rangle
\end{align}
 Moreover, the fact that the state created by the operator  $c{\partial X_{\mu}}$  is BRST exact and the identity
\begin{equation}\label{Uidentity}
U_nU_m^{\dagger}=U^{\dagger}_{2+\frac{2}{n}(m-2)}U_{2+\frac{2}{m}(n-2)}
\end{equation}
allow us to express \eqref{U3def} as follows
\begin{align}
&\left \langle c{\partial X_{\mu}}{\left({\sqrt{3}}\right)}c{\partial X_{\nu}}{\left(-{\sqrt{3}}\right)} U^{\dagger}_{\frac{8}{3}}\frac{b_0}{L_0} U_{\frac{8}{3}} {\partial X_{\sigma}}{\left(-\frac{1}{\sqrt{3}}\right)}c{\partial X_{\sigma}}{\left(\frac{1}{\sqrt{3}}\right)}\right\rangle \nonumber\\
&\qquad\qquad =\int_{0}^{\infty} dt \left \langle c{\partial X_{\mu}}{\left(-\frac{1}{a}\right)}{\partial X_{\nu}}{\left(\frac{1}{a}\right)}c {\partial X_{\sigma}}{\left(e^{-t}a\right)}c{\partial X_{\sigma}}{\left(-e^{-t}a\right)}\right\rangle
\end{align}
where $a=\sqrt{2}-1$. 

Combining all these ingredients reproduces the explicit expression for the quartic term in the Yang-Mills action  \cite{Berkovits:2003ny}.
\begin{align}\label{YMQ}
S^{(4)}=\frac{1}{4g^2}\int d^{p+1} x~ \text{Tr}\left(A_{\mu}A_{\nu}A_{\rho}A_{\sigma} \right)\left( S^{\mu\nu\rho\sigma}+T^{\mu\nu\rho\sigma}+U^{\mu\nu\rho\sigma} \right)
\end{align}
The terms $S^{\mu\nu\rho\sigma}, T^{\mu\nu\rho\sigma}, U^{\mu\nu\rho\sigma}$ are the contribtuions from $s, ~t,~u$-channel string field theory diagrams with only massive states propagating along the internal edge
 \begin{align}\label{smnu}
S^{\mu\nu\rho\sigma} &= \int_{0}^{\infty}dt~e^{-t} \left(e^{-2t}a^2-\frac{1}{a^2}\right)\left(\frac{e^{2t}}{48}\eta^{\mu\nu}\eta^{\rho\sigma}+\frac{1}{2}\frac{\eta^{\mu\sigma}\eta^{\nu\rho}}{\left( \frac{1}{a}-e^{-t}a\right)^4} \right)\nonumber\\
&=k_{e(r)}^{-2}D^{\mu\mu_{e(r)}\nu}_r\left(k_{e\left(r\right)}^2 D^{\rho\mu_{e(r)}\sigma}_r k_{e(r)}^{-2}\right)=\left(\eta \wedge \eta\right)^{\mu\nu\rho\sigma}
\end{align}
Here, the $e^{t}$ term in the expansion of the integrand arises due to the open string tachyon propgating along the internal propagator, for which the exponential representation of the propgator is invalid.  This can be dealt with by the analytic continuation
\begin{equation}\label{anc}
\int_{0}^{\infty} dt ~e^t =-1.
\end{equation}

\section{Inner product between squeezed states}
Identity for the inner product between two squeezed state of the bosonic $a$ oscillator \cite{Kostelecky:2000hz}
\begin{align}\label{MidentityG1}
\langle 0|\text{exp}\left(\lambda\cdot a+\frac{1}{2}a\cdot S\cdot a \right)&\text{exp}\left(\mu\cdot a^{\dagger}+\frac{1}{2}a^{\dagger}\cdot V\cdot a^{\dagger}\right)|0\rangle \nonumber\\
=&\text{Det}\left(1-S\cdot V \right)^{-1/2}\text{exp}\left(\lambda\cdot\left(1- V\cdot S\right)^{-1}\cdot \mu\right) \nonumber\\
& \text{exp}\left(\frac{1}{2}\lambda\cdot \left(1-V\cdot S \right)^{-1}\cdot V\cdot \lambda+\frac{1}{2}\mu\cdot \left( 1-S\cdot V\right)^{-1}\cdot S\cdot \mu\right) 
\end{align}
Similar identity exists  for the fermionic  $b, c$ oscillators also
\begin{align}\label{Midentitybc}
\langle 0| \text{exp}&\left(b\cdot \lambda_b-\lambda_c\cdot c -b\cdot  S\cdot c\right)\text{exp}\left(b^{\dagger}\cdot \mu^b+\mu^c\cdot c^{\dagger}+b^{\dagger}\cdot V\cdot c^{\dagger}\right)|0\rangle \nonumber\\
=&\text{Det}\left( 1- S\cdot  V\right)\text{exp}\left( -\lambda_c\cdot \left(1-V\cdot S \right)^{-1}\cdot \mu_b-\mu_c\cdot\left( 1-S\cdot V\right)^{-1}\cdot \lambda_b\right) \nonumber\\
& \text{exp}\left(\lambda_c\cdot\left( 1-V\cdot S\right)^{-1}\cdot V\cdot \lambda_b+\mu^c\cdot S\cdot\left( 1- S\cdot  V\right)^{-1}\cdot   \mu^b\right) \nonumber\\
\end{align}

\end{document}